\pdfoutput=1
\documentclass[doublecol]{epl2}

\usepackage{ucs,graphicx,subfigure}
\usepackage[utf8x]{inputenc}
\graphicspath{{./figures/}}

\begin{document}

\title{Magnetic properties of La$_{0.67}$Sr$_{0.33}$MnO$_3$/BiFeO$_3$(001) heterojunctions: chemically abrupt versus atomic intermixed interface}
\shorttitle{Magnetic properties of the LSMO/BFO(001) heterojunction}

\author{R. F. Neumann\inst{1,2} \and M. Bahiana\inst{1} \and N. Binggeli\inst{2,3}}
\shortauthor{R. F. Neumann \etal}

\institute{
  \inst{1} Instituto de F\'isica, Universidade Federal do Rio de Janeiro, Caixa Postal 68528, Rio de Janeiro 21941-972, Brazil \\
  \inst{2} Abdus Salam International Centre for Theoretical Physics, Trieste 34014, Italy \\
  \inst{3} IOM-CNR DEMOCRITOS National Simulation Center, Trieste 34014, Italy \\
}


\abstract{Using first-principles density-functional calculations, we address the magnetic properties of the ferromagnet/antiferromagnet La$_{0.67}$Sr$_{0.33}$MnO$_3$/BiFeO$_3$(001) heterojunctions, and investigate possible driving mechanisms for a ferromagnetic (FM) interfacial ordering of the Fe spins recently observed experimentally. We find that the chemically abrupt defect-free La$_{0.67}$Sr$_{0.33}$MnO$_3$/BiFeO$_3$(001) heterojunction displays, as ground state, an ordering with compensated Fe spins. Cation Fe/Mn intermixing at the interface tends to favour, instead, a FM interfacial order of the Fe spins, coupled antiferromagnetically to the bulk La$_{0.67}$Sr$_{0.33}$MnO$_3$ spins, as observed experimentally. Such trends are understood based on a model description of the energetics of the exchange interactions.}

\maketitle

\section{Introduction}

Current research on multiferroics is creating exciting possibilities for electric-field control of magnetisation (see, e.g., Refs.  \cite{ramesh2007multiferroics,bibes2008spintronics,wu2010reversible,chu2008fmcontrol}). Such control has potential applications for novel data storage, spintronics, and high-frequency magnetic devices. One of the few single-phase room-temperature multiferroic candidate is BiFeO$_3$ (BFO), which is a ferroelectric (FE) ($T_C \sim 1103$ K) antiferromagnetic (AFM) ($T_N \sim 643$ K) insulator \cite{wang2003epitaxial}. Unlike most FE ferromagnetic (FM) single-phase multiferroics, BFO is characterised by a robust coupling between its FE and AFM ordering, which enables electrical control of AFM domains in BFO films at room temperature \cite{zhao2006afcontrol}.

Being an antiferromagnet, however, BFO needs to be coupled to a FM layer by an exchange bias (EB) mechanism, in order to couple and possibly manipulate the latter magnetisation with the BFO polarisation.  Recently, in fact, electrical control of EB \cite{wu2010reversible} and of local ferromagnetism \cite{chu2008fmcontrol} have been achieved in some ferromagnet/BFO  bilayers. In particular, reversible switching by an electric field between two EB states has been demonstrated by Wu {\it et al.} \cite{wu2010reversible} in La$_{0.67}$Sr$_{0.33}$MnO$_3$/BFO(001) bilayers composed of a few-nanometers-thick layer of La$_{0.67}$Sr$_{0.33}$MnO$_3$ (LSMO), a FM ($T_C \sim 370$ K) half-metal \cite{krishnan2011microstructural}, on a BFO(001) thin film.

In such LSMO/BFO(001) bilayers, an intriguing FM ordering of the Fe atoms at the interface has been observed by Yu  {\it et al.} \cite{yu2010interface} in x-ray circular dichroism experiments. The corresponding Fe FM moment was found to be coupled antiferromagnetically to the LSMO FM layer and to give rise to a significant EB \cite{yu2010interface}. The cause of this Fe FM ordering is actively discussed, but is not yet a resolved issue, and is important for potential optimisation.

Several candidate microscopic mechanisms have been invoked. The FM state was initially ascribed to a possible strong Mn-Fe hybridisation across the interface \cite{yu2010interface}, related to orbital ordering. Subsequently, the FM state was also suggested to be due to alteration of the Fe-O-Fe angles and suppression of octahedral tilting in the vicinity of the interface \cite{borisevich2010suppression}. Very recently, model Hamiltonian studies indicated, instead, that the Fe FM moment could be due to the doping generated by a BFO positive FE surface charge in the BFO unit layer adjacent to the LSMO, which in turn could induce ferromagnetism in this region  \cite{calderon2011magnetoelectric}.

Here we use {\it ab initio} density-functional-theory (DFT) calculations to investigate the magnetic properties of LSMO/BFO(001) heterojunctions and probe potential microscopic mechanisms which may lead to Fe ferromagnetism at the interface. {\it Ab initio}  calculations are well suited, in general, to examine microscopic effects, such as hybridisation, atomic-structure, and FE-charge effects on the local magnetic structure of interfaces. They could provide useful complementary information to identify the origin of the interfacial Fe FM ordering.  To the best of our knowledge, however, only one study reported very recently {\it ab initio} results for the FE properties of the chemically abrupt LSMO/BFO(001) heterojunction\cite{yu2012control}, and no first-principles study of the magnetic properties of this system has been reported so far.

On the basis of the {\it ab initio} calculations, we will show in particular that, whereas Fe FM ordering is not energetically favoured at the chemically abrupt defect-free LSMO/BFO(001) heterojunctions, atomic interdiffusion including Fe/Mn intermixing can generate an Fe FM moment at the interface coupled antiferromagnetically to that of the LSMO.  Such a behaviour is understood more generally based on a model of the energetics of the exchange interactions. Our first-principles calculations also reveal a switchable Fe FM state at the intermixed interfaces, which depends on the sign of the BFO FE polarisation and may account for the reported electric-field control of the EB at the LSMO/BFO interface.
\section{Computational details}

The DFT  calculations were performed with the {\sc Quantum-ESPRESSO} \cite{QE-2009} package, which uses pseudopotentials and a plane-wave basis set. The computations were carried out within the generalised gradient approximation (GGA) with the spin-polarised Perdew-Burke-Ernzerhof exchange-correlation functional. We employed scalar-relativistic ultrasoft pseudopotentials generated using the following reference atomic configurations: $5d^{10} 6s^2 6p^3$ for Bi, $3d^7 4s^1 4p^0$ for Fe, $5s^2 5p^6 5d^1 6s^{1.5} 6p^{0.5}$ for La, $4s^2 4p^6 4d^1 5s^1 5p^0$ for Sr, $3s^2 3p^6 3d^{4.5} 4s^{0.4} 4p^{0.1}$ for Mn and $2s^2 2p^4$ for O. The nonlinear-core correction to the exchange-correlation potential was included for Fe, La, Sr and Mn. To represent the La/Sr alloying in LSMO, we have used the virtual crystal approximation, replacing the La and Sr atoms by a virtual ``LS'' atom, whose pseudopotential is a La$_{0.67}$Sr$_{0.33}$ weighted average of the original pseudopotentials. The LSMO/BFO heterojunctions
were modelled using slab geometries in periodically repeated supercells.

A kinetic energy cutoff of 30 Ry (300 Ry) was used for the plane-wave expansion of the wavefunctions (electronic density). For the heterojunctions, we employed supercells including 100 atoms, namely 10 perovskite ABO$_3$(001) atomic bilayers along the growth direction and using laterally a reconstructed $(\sqrt{2} \times \sqrt{2}) R 45^\circ$ perovskite surface unit cell to accommodate the BFO's $G$-type antiferromagnetism. The self-consistent supercell calculations were performed using a $6 \times 6 \times 1$ $\Gamma$-centred Monkhorst–Pack grid and a 10 mRy electronic level smearing for the Brillouin zone integrations.  Unless otherwise specified, all atomic positions were relaxed until the magnitude of the residual force acting on each atom was smaller than 5 mRy/au.

We considered LSMO pseudomorphically grown on BFO,  as in the experimental study \cite{yu2010interface}. To construct the supercells, we used the theoretical lattice parameter values obtained from separate bulk calculations for BFO and LSMO, using $8 \times 8 \times 8$ and $12 \times 12 \times 12$ {\bf k}-point grids, respectively.  For BFO, a 10-atom rhombohedral cell was used, with the rhombohedral angle $\alpha$ set to $60^\circ$ \cite{neaton2005bulkBFO}. The following values were obtained for the equilibrium pseudocubic lattice parameter: $a = 5.64/\sqrt{2} \mbox{\ \AA} = 3.99 \mbox{\ \AA}$ and Wyckoff positions ($R3c$ space group): $w_{\mbox{\tiny Bi}} = 0$, $u_{\mbox{\tiny Fe}} = 0.225$, $x_{\mbox{\tiny O}} = 0.536$, $y_{\mbox{\tiny O}} = 0.936$ and $z_{\mbox{\tiny O}} = 0.387$. These values are in good agreement with experiment and previous DFT values \cite{neaton2005bulkBFO}, and lead to FeO$_6$ octahedra tilted by $11.8^\circ$ with respect to the [001]
axis ($a^{-}a^{-}a^{-}$ tilt system in Glazer notation). To estimate the out-of-plane $c/a$ ratio of LSMO grown on top of BFO (under tensile inplane strain),\footnotemark[1] we used a 5-atom tetragonal unit cell \cite{vailionis2011misfit} with in-plane lattice constant $a$, and varied $c/a$. The resulting LSMO tetragonal distortion obtained from total-energy minimisation was $c/a = 0.958$.  The supercell size was then taken as $a \sqrt{2} \times a \sqrt{2} \times L$, where $L = \left(a n_{\mbox{\tiny BiO}} + c n_{\mbox{\tiny LSO}} \right)$ and $n_{\mbox{\tiny BiO}}$ ($n_{\mbox{\tiny LSO}}$) is the number of Bi (LS) oxide layers.

\footnotetext[1]{The theoretical inplane lattice parameter mismatch between LSMO and BFO is 2.6 \%, similar to the experimental misfit value of 2.3 \% \cite{krishnan2011microstructural}}

\begin{figure}[ht]
\centering
\resizebox{0.12\textwidth}{!}{\includegraphics{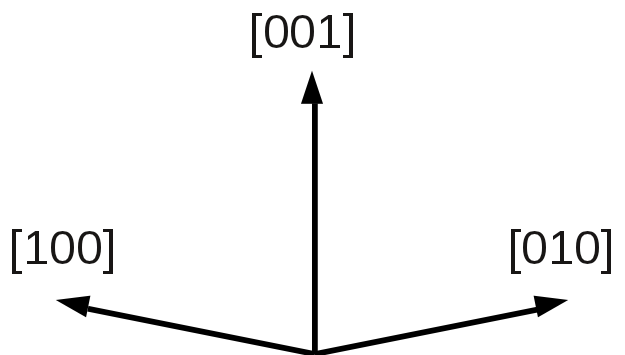}}
\resizebox{2.2cm}{!}{\includegraphics{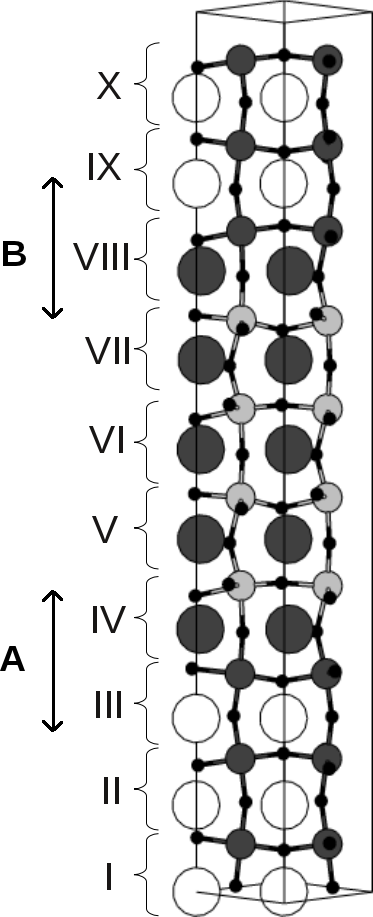}}
\resizebox{2.2cm}{!}{\includegraphics{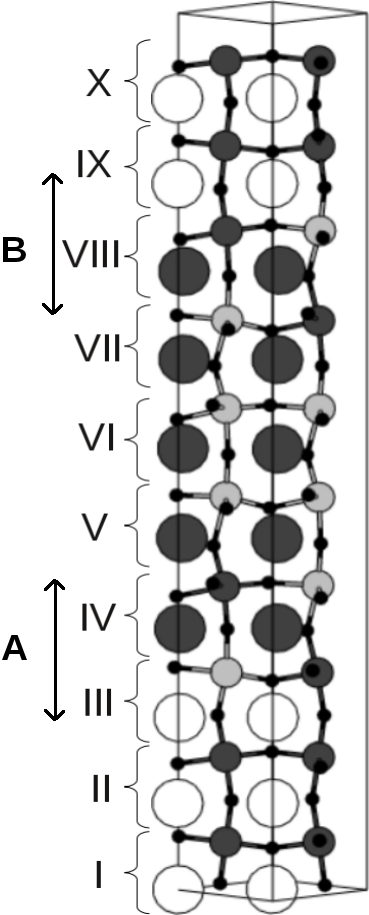}}
\resizebox{0.04\textwidth}{!}{\includegraphics{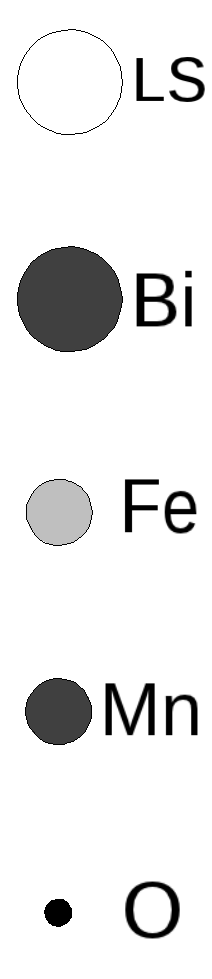}}
\caption{Supercells used to model the LSMO/BFO(001) heterojunctions with the chemically abrupt MnO$_2$-terminated LSMO (001) interfaces (left) and the corresponding Fe/Mn intermixed interfaces (right). The thin solid line indicates the supercell and the rods show the Fe(Mn)-O bonds. The atomic positions shown are for the fully relaxed internal atomic structure. }
\label{snapshot}
\end{figure}

We considered both terminations for the chemically abrupt interface, namely the MnO$_2$-terminated LSMO (001) interface (with atomic layer stacking: -LSO-MnO$_2$-BiO-FeO$_2$- ) and the LSO-terminated interface (with stacking: -MnO$_2$-LSO-FeO$_2$-BiO- ). In both cases we investigated the effect of Fe/Mn intermixing at the interface considering two intermixed Fe$_{0.5}$Mn$_{0.5}$ atomic layers at the junction. The supercells used for the abrupt MnO$_2$-terminated interface and for the corresponding Fe/Mn intermixed interface are displayed in fig. \ref{snapshot}. Each supercell contains two interfaces with the same chemical termination (A and B interfaces in fig. \ref{snapshot}), which would be equivalent without the BFO FE distortion. In this paper, we will concentrate mostly on the MnO$_2$-terminated interface and on the corresponding intermixed interfaces, and only briefly mention the results for the LSO-related interfaces, as they are very similar and lead to the same conclusions.

The internal atomic positions in the supercells were relaxed starting from a geometry with $6^\circ$ tilted FeO$_6$ octahedra (having the same symmetry as in the bulk, but with a tilting distortion of smaller amplitude) in the central part of the BFO slab (region V-VI in fig. \ref{snapshot}) and non-tilted MnO$_6$ octahedra in the LSMO slab. The resulting BFO distortion yields a FE polarisation in BFO pointing from the A to the B interface (with negative and positive BFO FE surface charge, respectively).\footnotemark[2] The starting tilt pattern was introduced just to break the symmetry of the (non-tilted) paraelectric phase  \cite{junquera2003critical} and let the atomic structure relax to the lowest-energy configuration. The octahedral tilts in the final relaxed geometry were calculated as the inclination angles of the line joining the bottom and top oxygens of each octahedron with respect to the growth axis (taken as the vertical axis). The relaxation calculations were performed starting from
different magnetic configurations for the Fe monolayer (ML) closest to the interface, corresponding to heterojunctions with compensated or uncompensated Fe spins.

\footnotetext[2]{Compared to Ref. \cite{yu2012control}, in our case 
the two interfaces in the supercell are chemically equivalent and only the FE 
contribution $- \Delta_1 /2$ (at the A interface) and $+\Delta_1 /2$ (at the B 
interface) in fig. 4 (b) of Ref.  \cite{yu2012control} make the two interfaces 
inequivalent. }
\section{Ab-initio results}

\begin{figure}[ht]
\centering
\subfigure[abrupt_bulk][Type-I: Fe-AFM ML]{\includegraphics[width=0.22\textwidth]{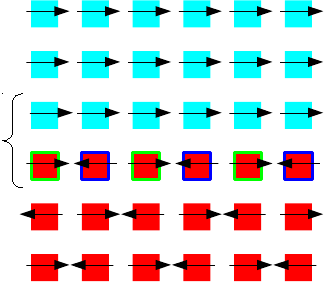}}
\hfill
\subfigure[abrupt_AF][Type-II: Fe-FM ML]{\includegraphics[width=0.22\textwidth]{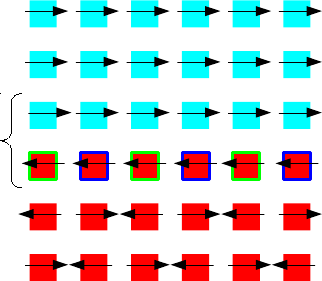}} \\
\caption{Schematic representations of the configurations with compensated (a) and uncompensated (b) Fe spins at the chemically abrupt LSMO/BFO(001) interface. Cyan (light gray) and red (dark gray) squares indicate Mn and Fe atoms, respectively. Type-I: an AFM-Fe ML is present at the interface, as in bulk BFO. Type-II: a FM-Fe ML is present with a moment antiparallel to that of bulk LSMO. }
\label{Jn_model_abrupt}
\end{figure}

We examined the relative stability of the configurations with compensated (type-I) and uncompensated (type-II) Fe spins schematically represented in fig.  \ref{Jn_model_abrupt}, at the chemically abrupt A and B interfaces of LSMO/BFO(001). The type-I configuration corresponds to an AFM Fe ML at the junction, as in bulk BFO, and type-II to a FM Fe ML at the interface with its magnetic moment aligned antiparallel to the FM moment of bulk LSMO (as suggested experimentally \cite{yu2010interface}). For both the MnO$_2$- and LSO- terminated interfaces, we find that the lowest-energy magnetic state is, by far, the compensated type-I configuration, and this is independent of the BFO polarisation direction, i.e., both for the A and B interfaces. Type-I is lower in energy by as much as 0.52 (0.38) eV per swapped Fe spin than type-II at the A (B) MnO$_2$-terminated interface, and by 0.39 (0.50) eV per swapped Fe spin at the A (B) LSO-terminated interface.

\begin{figure}[ht]
\centering
\includegraphics[width=0.49\textwidth]{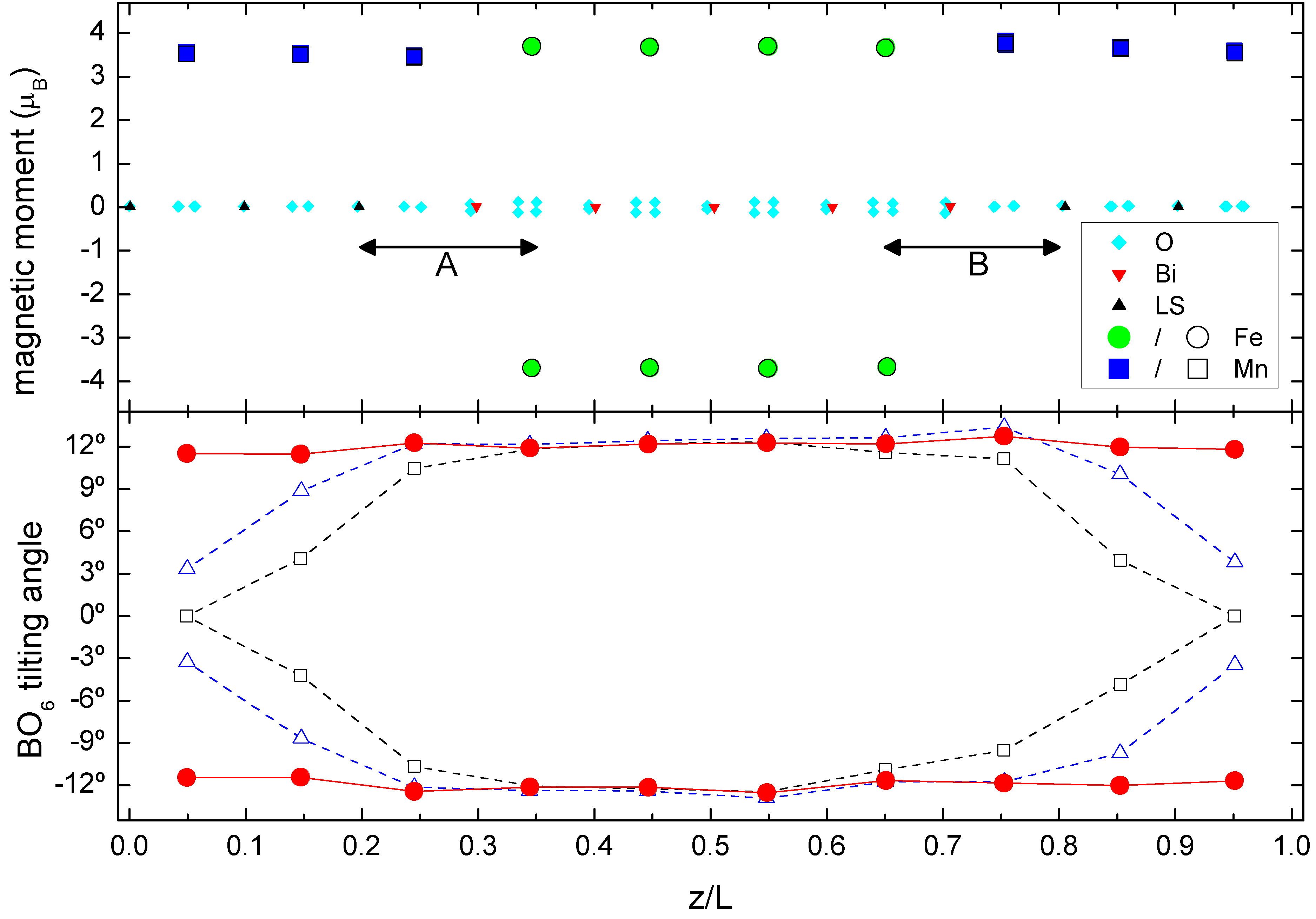}
\caption{Atomic resolved magnetic moments (upper panel) and octahedral tilt angles (lower panel) in the LSMO/BFO/LSMO heterostructure with chemically abrupt MnO$_2$-terminated interfaces. Positive (negative)  tilt angles refer to clockwise (anticlockwise) rotations. The A and B interface regions (see fig. \ref{snapshot}) are also indicated. Filled symbols are for the fully relaxed internal atomic structure. Open triangles (squares) in the lower panel show the tilt angles from relaxation calculations with the 3 central LSMO atomic layers (3 LSMO bilayers) kept frozen in the untilted cubic perovskite geometry. Open circles (squares) in the upper panel show the Fe (Mn) moments from calculations with the 3 frozen LSMO bilayers.}
\label{Mn_mag_struct}
\end{figure}

In fig. \ref{Mn_mag_struct}, we display, as a function of the layer position, the atomic-resolved magnetic moments (upper panel) and octahedral tilt angles (lower panel, filled  symbols) for the magnetic ground state of the LSMO/BFO/LSMO heterostructure with chemically abrupt MnO$_2$-terminated interfaces. The Fe (Mn) magnetic moments retain their bulk value of 3.68 $\mu_{\mbox{\tiny B}}$ (3.67 $\mu_{\mbox{\tiny B}}$), to within 0.02 $\mu_{\mbox{\tiny B}}$ (0.15 $\mu_{\mbox{\tiny B}}$), in all FeO$_2$ (MnO$_2$) layers, including those closest to the junction. For both types of termination, we find that, over the entire BFO slab, the FeO$_6$ octahedra exhibit tilt-angle values very close to the bulk value ($11.8^\circ$). The MnO$_6$ octahedra are also highly tilted, with angles as large as $\sim12^\circ$ at the interface, which decrease only very slowly within the LSMO slab towards their bulk strained value ($9.5^\circ$). The slow tilting decay within LSMO is confirmed by calculations we performed using a
larger (120-atom) supercell with 2 additional LSMO unit layers. In this case, the tilt angle further decreases to $10.6^\circ$ in the central part of the LSMO slab, with no significant change in the results for the other layers displayed in fig. \ref{Mn_mag_struct}.

To check the robustness of the magnetic ground state, we also carried out atomic relaxation calculations in which we keep  frozen (in the initial untilted geometry) either just the three central atomic layers in LSMO or the three bilayers X, I, and II in LSMO (see fig. \ref{snapshot}); the latter boundary conditions would correspond to a rigid non-tilted (simple cubic) perovskite substrate close to the LSMO/BFO interface. In both cases, the magnetic ground state is the type-I configuration, with similar energy differences (to within 0.05 eV) compared to the relaxed case. The corresponding Fe and Mn magnetic moments and octahedral tilt angles are also displayed in fig. \ref{Mn_mag_struct} (open symbols). The Fe and Mn magnetic moments remain virtually unchanged with respect to the relaxed case. We also note that, even with rigid (zero tilt) boundary condition on the three LSMO bilayers, all FeO$_6$ octahedra in BFO recover their bulk tilting value (to within $1^\circ$) and the MnO$_6$ octahedra at the
interface are in a similar highly tilted configuration.

The behaviour we find for the octahedral tilts in fig. \ref{Mn_mag_struct} can be understood based on the energetics of the tilting distortions in BFO and LSMO. From separate bulk calculations, we obtain that the energy lowering from the non-tilted to the bulk-tilted ground-state configuration is 1.2 eV/f.u. for BFO,\footnotemark[3] whereas it is only 0.1 eV/f.u. for strained LSMO. The BFO octahedral tilting is thus extremely robust compared to that of LSMO. The strong BFO tilting is therefore expected to dominate near the interface, and to impose, by continuity, an enhanced tilting in LSMO at the junction. The weak resistance of the LSMO octahedral network to this distortion and the deep extension of the deformation into the LSMO are consistent with the results of previous {\it ab initio} calculations for the octahedral tilts in La$_{0.75}$Sr$_{0.25}$MnO$_3$(001) slabs imposing untilted boundary conditions \cite{he2010control}.

\footnotetext[3]{The much stronger bulk FE distortion of BFO compared to that of BaTiO$_3$ ($\sim 0.02$ eV/f.u.)(Ref. \cite{junquera2003critical} and Ref. [22] therein) is also the reason we recover the bulk FE distortion even in a BFO film of small thickness, unlike for the BaTiO$_3$ films in Ref. \cite{junquera2003critical}. }

We note that the tensile strain tends to increase the LSMO tilt angle, from a value of $8.6^\circ$ for unstrained bulk LSMO to $9.5^\circ$ for LSMO strained to the BFO lattice constant. This trend is consistent with the one reported from EXAFS measurements \cite{vailionis2011misfit} and from previous GGA calculations for La$_{0.75}$Sr$_{0.25}$MnO$_3$ considering a smaller tensile strain \cite{he2010control}. Our tilt angle for unstrained LSMO also agrees, to within $0.8^\circ$, with the latter GGA value for the unstrained case. One may also note that the Mn magnetic moment, in fig. \ref{Mn_mag_struct}, is remarkably insensitive to the MnO$_6$ tilt angle. This is related to the presence of a robust bandgap in the LSMO minority spin channel, between occupied O 2p states and unoccupied Mn 3d states \cite{he2010control}.

Hence, for the chemically abrupt defect-free LSMO/BFO(001) heterojunctions pseudomorphically strained to BFO, our calculations yield as robust ground state the compensated type-I configuration (no Fe ferromagnetism). Experimentally, however, it is known that for other related epitaxial oxide interfaces, such as the LSMO/SrTiO$_3$(001) heterojunctions, cation interdiffusion does occur over a few unit-cell layers across the interface \cite{pailloux2002nanoscale,herger2008structure}. Also, very recently, atomic intermixing at the interface was reported in LSMO/BFO(001) epitaxial heterojunctions \cite{krishnan2011microstructural}. This motivated us to investigate whether atomic intermixing could possibly favour Fe ferromagnetism at the interface.

\begin{figure}[ht]
\centering
\includegraphics[width=0.44\textwidth]{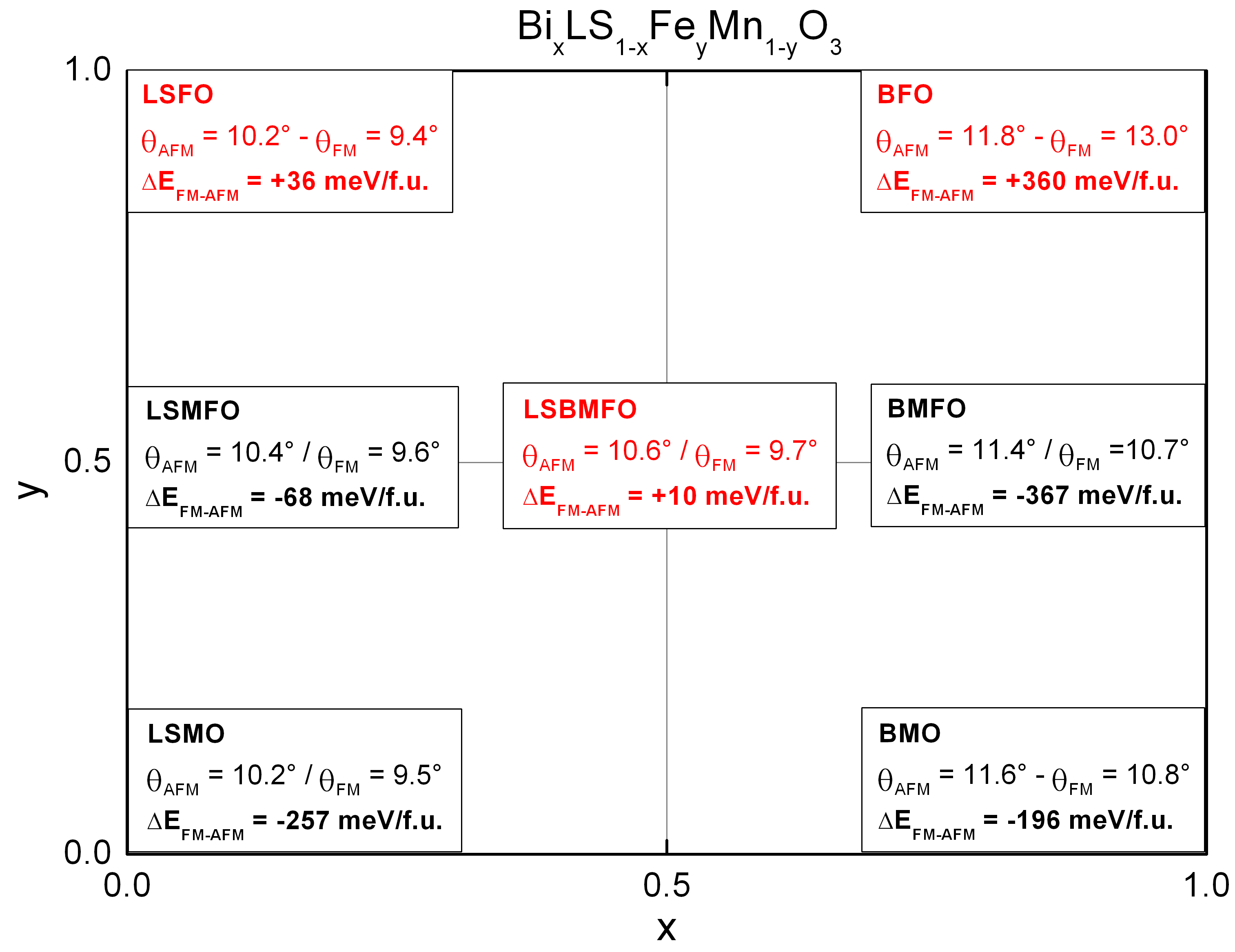}
\caption{Diagram indicating the magnetic ground state of the Bi$_x$LS$_{(1-x)}$Fe$_y$Mn$_{(1-y)}$O$_3$ alloy for selected compositions ($x,y$), as obtained from bulk calculations, with the BFO lattice constant, in a 10-atom rhombohedral unit cell. ${\Delta E}_{\mbox{\tiny FM-AFM}}$ is the energy difference between the FM and the AFM phase and $\theta_{\mbox{\tiny AFM (FM)}}$ is the tilt angle in the AFM (FM) phase.}
\label{alloy_diagram}
\end{figure}

In order to guide our search for the type of intermixing which may favour an AFM alignment of the Fe relative to the Mn spins and/or Fe ferromagnetism, we first examined the magnetic properties of bulk Bi$_x$LS$_{1-x}$Fe$_y$Mn$_{1-y}$O$_3$ alloys for selected $(x,y)$ compositions. The calculations, including internal atomic relaxation, were performed using a 10-atom rhombohedral cell with the same computational parameters as for bulk BFO. We used the BFO substrate lattice parameter $a$, and considered the FM and AFM configurations for the two B-site cations in the unit cell. In fig. \ref{alloy_diagram}, we report the resulting magnetic ground state for each of the investigated alloys, as well as the energy difference, ${\Delta E}_{\mbox{\tiny FM-AFM}}$, between the FM and AFM phase and the tilt angle, $\theta_{\mbox{\tiny AFM (FM)}}$, of the AFM (FM) state.

The results in fig. \ref{alloy_diagram} show that, in general, increasing the Fe content favours the AFM alignment (${\Delta E}_{\mbox{\tiny FM-AFM}}>0$ for LSFO and BFO). However, in the presence of Mn, it is for $(0.5,0.5)$ that we find a potential candidate leading to an AFM alignment of the Fe and Mn spins. The LS$_{0.5}$Bi$_{0.5}$Mn$_{0.5}$Fe$_{0.5}$O$_3$ ground state is AFM, with an energy ${\Delta E}_{\mbox{\tiny FM-AFM}}$ of 10 meV per Fe to reverse the Fe spins.

\begin{figure}[ht]
\centering
\subfigure[mixed_bulk][Type I: Fe-AFM ML]{\includegraphics[width=0.22\textwidth]{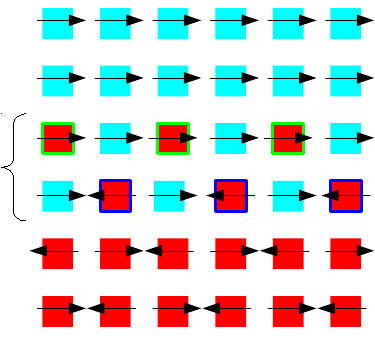}}
\hfill
\subfigure[mixed_AF][Type II: Fe-FM ML]{\includegraphics[width=0.22\textwidth]{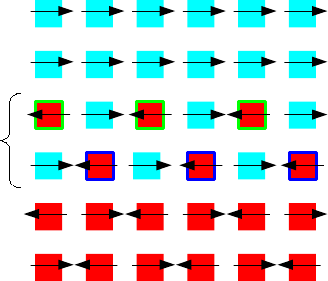}}
\caption{Same schematic representations of the compensated (a) and uncompensated (b) Fe spins configurations as in fig. \ref{Jn_model_abrupt}, but for the Fe/Mn intermixed LSMO/BFO(001) interface. }
\label{Jn_model_intermixed}
\end{figure}

\begin{figure}[ht]
\centering
\includegraphics[width=0.49\textwidth]{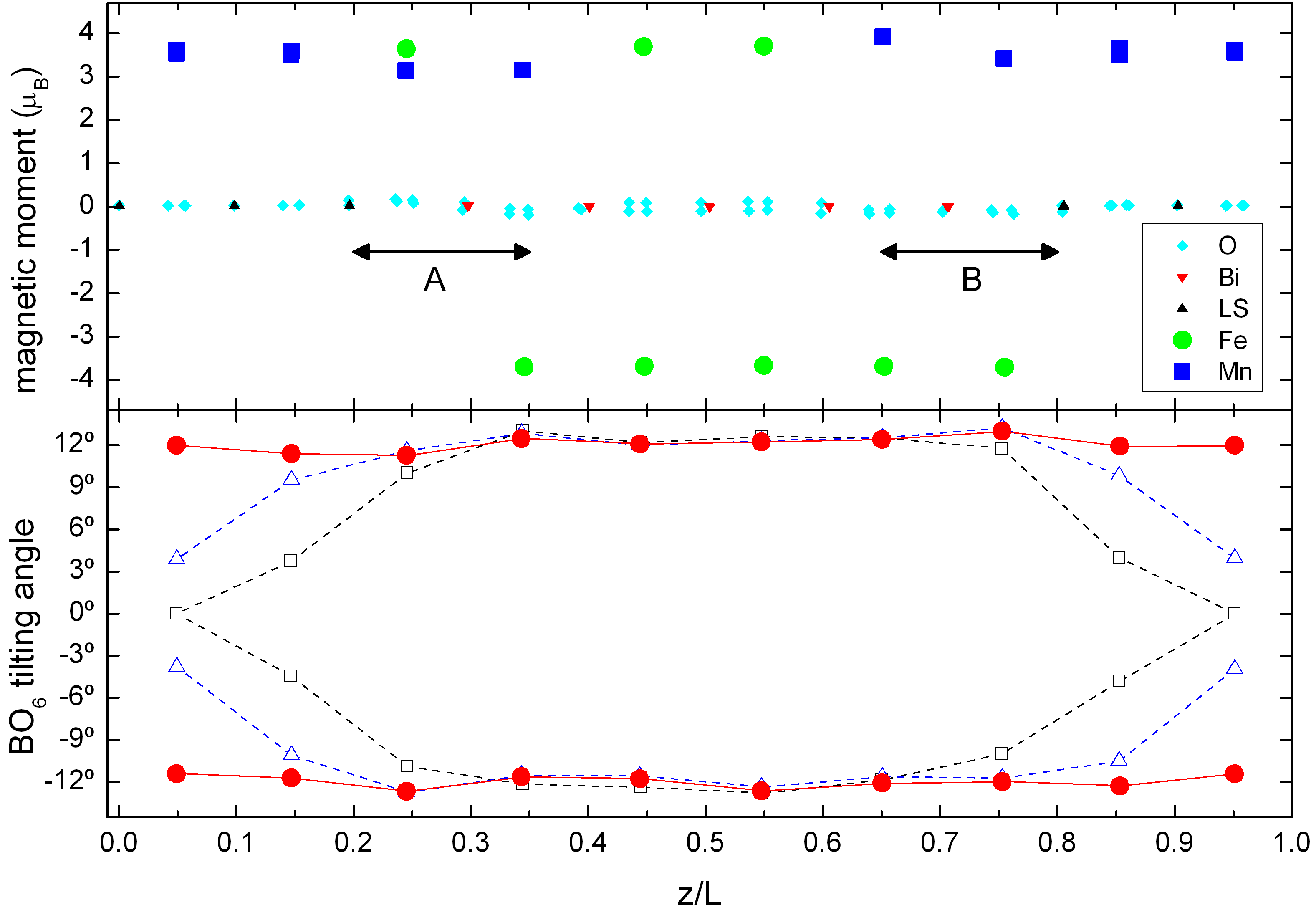}
\caption{Same data as in fig. \ref{Mn_mag_struct}, but for the LSMO/BFO/LSMO heterostructure with Fe/Mn intermixed interfaces.}
\label{Mn_imix_mag_struct}
\end{figure}

Encouraged by the bulk-alloy results, we have then considered an ultrathin alloy layer at the LSMO/BFO interface with 50\% of Fe and 50\% of Mn (see fig. \ref{Jn_model_intermixed}). The latter was obtained by exchanging, at the abrupt interface, Fe and Mn atoms from the FeO$_2$ and MnO$_2$ monolayer closest to the junction. This also leads to a  50\%-50\%  Bi-LS local environment for one of the mixed Fe$_{0.5}$Mn$_{0.5}$ monolayer (see fig. \ref{snapshot}). The resulting ground-state magnetic structure and tilt angles obtained for the heterostructure with MnO$_2$-terminated interfaces are displayed in fig. \ref{Mn_imix_mag_struct}.

Both for the MnO$_2$- and LSO-terminated junctions, we find that the lowest energy magnetic configuration at the B interface changes from type-I (compensated Fe spins) to type-II (FM Fe) at the Fe/Mn intermixed junction. At the B interface, the type-II configuration, in fig. \ref{Jn_model_intermixed}b, is 85 (82) meV lower in energy per swapped Fe spin than the type-I configuration, in fig. \ref{Jn_model_intermixed}a, for the MnO$_2$- (LSO-) related termination.  At the A interface, instead, type-I remains the ground state, although the energy difference between type II and type I decreases drastically, i.e., from 0.52 (0.39) eV to 0.08 (0.07) eV with intermixing at the  MnO$_2$- (LSO-) terminated junction. We note that with the constraints of frozen LSMO layers, the relative stability of the type-II configuration at the B interface remains essentially unchanged (86 meV per swapped Fe spin with the three frozen bilayers and MnO$_2$-related termination).

Inspection of fig. \ref{Mn_imix_mag_struct} indicates that the magnitude of the Fe magnetic moments in the interface region remains similar to that in bulk BFO (to within 0.02 $\mu_{\mbox{\tiny B}}$). We observe, instead, a slightly larger variation in the Mn moment, which ranges from 3.0 $\mu_{\mbox{\tiny B}}$ (A interface) to 3.9 $\mu_{\mbox{\tiny B}}$ (at the interface B). Regarding the octahedral tilt angles, the introduction of intermixing does not alter significantly the results compared to those at the abrupt interface.

{Going back to the experimental situation, when no electric pulse is applied to polarise the BFO (as in the experiment by Yu {\it et al.} \cite{yu2010interface}) one would expect domains of both A and B type to be present at the junction. In this case, based on our results, the Fe/Mn intermixing is expected to lead to uncompensated Fe spins coming from the B domains. We note, in this connection, that in our supercell calculations we have used a relatively large (50~\%) Fe/Mn intermixing at the interface (which makes the computations tractable). The results, however, on the stability of the uncompensated Fe spins at the B interface are expected to hold also locally with only very few intermixed Fe/Mn atoms at the interface, in view of the short-range nature of the (super-)exchange interactions. Possibly, in this case, the stabilisation energy per swapped (uncompensated) Fe spin of intermixed Fe may be somewhat reduced compared to our very hefty value of $\sim 80$ meV.

When an electric pulse is applied to polarise the BFO (as in the experiment by Wu {\it et al.} \cite{wu2010reversible}) and used to switch then between negative and positive BFO FE polarisation (A and B interface, respectively), in one case (B) one expects the configuration with uncompensated Fe spins to be established (type II), and in the other case (A) that with compensated Fe spins (type I). This is consistent with an EB, produced by the uncompensated Fe spins, switched off and on when the BFO polarisation is flipped, as is observed experimentally \cite{wu2010reversible}. The latter magnetoelectric effect is in agreement with the mechanism suggested by the model Hamiltonian study in Ref.  \cite{calderon2011magnetoelectric}. Our results, however, indicate in this connection that the uncompensated Fe spins and the associated magnetoelectric switching effect occur only when Fe/Mn intermixing is introduced.

Our {\it ab initio} results show thus that uncompensated Fe spins can emerge at the LSMO/BFO interface as a consequence of Fe/Mn intermixing. The physical origin of this behaviour can be understood based on a model description ($J_n$-model) considering only the exchange interactions between nearest-neighbour B-site cations. The model shows that intermixing releases the constraints on the Mn-Fe exchange coupling for the occurrence of the Fe FM state at the interface and extends its stability range to the physical range of weak Fe-Mn AFM interactions.

\section{$J_n$-model interpretation}

Within this model, the LSMO/BFO(001) interface is viewed as a cubic lattice of Mn and Fe atoms with Ising-like spins (see figs. \ref{Jn_model_abrupt} and \ref{Jn_model_intermixed}). We consider all Mn spins as fixed in the FM ordering and the Fe spins located outside the bilayer interface region, highlighted in figs. \ref{Jn_model_abrupt} and \ref{Jn_model_intermixed}, as fixed in the bulk G-type AFM ordering. For a given Fe/Mn atomic arrangement, either abrupt (fig. \ref{Jn_model_abrupt}) or intermixed (fig. \ref{Jn_model_intermixed}), the exchange energy of the possible spin configurations can be calculated in terms of the exchange coupling constants as a function of the orientations of the free Fe interface spins, $S_1 = \pm 1$ and $S_2 = \pm 1$ [whose symbols are surrounded by light green (gray) and dark blue (black) lines, respectively]. The exchange constants that are relevant to this calculation are $J_{\mbox{\scriptsize Fe-Fe}} < 0$ (the Fe-Fe interaction inside BFO) and $J_{\mbox{\scriptsize Fe-Mn}}
$ (the Fe-Mn interaction across the interface).\footnotemark[4]

\footnotetext[4]{For simplicity, we used a single set of Fe-Mn and Fe-Fe exchange parameters. Using two sets of parameters (one for intra-bilayer coupling and another for inter-bilayer coupling, with $J^{\mbox{\tiny Intra}}_{\mbox{\tiny Fe-Mn}} \neq J^{\mbox{\tiny Inter}}_{\mbox{\tiny Fe-Mn}}$ and $J^{\mbox{\tiny Intra}}_{\mbox{\tiny Fe-Fe}} \neq J^{\mbox{\tiny Inter}}_{\mbox{\tiny Fe-Fe}}$) yields the same type of relaxation on the conditions for the emergence of a type-II ground state.}

For the abrupt interface (fig. \ref{Jn_model_abrupt}), the exchange energy per $S_1$ and $S_2$ spin is: $E(S_1,S_2) = J_{\mbox{\scriptsize Fe-Fe}} (S_1 - S_2 - 7 S_1 S_2) - J_{\mbox{\scriptsize Fe-Mn}} (S_1 + S_2)$. In order to have the experimentally observed magnetic configuration as the lowest in energy, we need to have $E_{\mbox{\tiny type II}} = E(-1,-1)$ (fig. \ref{Jn_model_abrupt}b) lower in energy than $E_{\mbox{\tiny type I}} = E(+1,-1)$ (fig. \ref{Jn_model_abrupt}a) and than $E(\pm 1,+1)$ (not shown). This is possible only if $J_{\mbox{\scriptsize Fe-Mn}}<0$ and $|J_{\mbox{\scriptsize Fe-Mn}}|>8|J_{\mbox{\scriptsize Fe-Fe}}|$. The exchange coupling constant $J_{\mbox{\scriptsize Fe-Fe}}$ can be estimated as the AFM Fe-Fe coupling present in bulk BFO, which according to the data in fig. \ref{alloy_diagram} is about -30 meV. This implies that the Fe-Mn coupling $J_{\mbox{\scriptsize Fe-Mn}}$ would have to be, not only AFM, but of the order of -240 meV to stabilise the observed
magnetic order, while typical values of exchange coupling constants are of the order of a few dozens of meV.

At the intermixed interface (fig. \ref{Jn_model_intermixed}), the exchange energy per $S_1$ and $S_2$ spin reads $E(S_1,S_2) = -J_{\mbox{\scriptsize Fe-Fe}} S_2 - J_{\mbox{\scriptsize Fe-Mn}} (6 S_1 + 5 S_2)$. In this case, it is sufficient to have just $J_{\mbox{\scriptsize Fe-Mn}}<0$ (AFM coupling) in order to have $E_{\mbox{\tiny Type II}}$ as the lowest possible energy state. The $J_n$-model shows thus that the requirement on the exchange constants for stabilising the type II configuration as ground state is much easier to fulfil when there is Fe-Mn intermixing at the interface, in correspondence with the increased number of Mn first-nearest neighbours each Fe can interact with.

\section{Conclusions}

We have investigated the magnetic properties of LSMO/BFO(001) heterojunctions, including chemically abrupt as well as atomic intermixed interfaces, using {\it ab initio} DFT calculations.  We were interested, in particular, in the possible reasons for the experimentally observed Fe FM ordering at the interface with Fe moment aligned anti-parallel to that of the LSMO.

We find that the chemically abrupt defect-free LSMO/BFO(001) interfaces display, as ground state, a configuration with compensated Fe spins (no Fe ferromagnetism). This holds both for the MnO$_2$ and LSO terminations of the LSMO(001) at the junction and is valid independent of the BFO polarisation, i.e., both at the interfaces with positive and negative BFO FE surface charge.

Interfacial Fe/Mn atomic intermixing tends to energetically favour, instead, an Fe FM ordering at the interface, with the Fe moment coupled antiferromagnetically to that of the LSMO. In particular, we find such a ground state, with uncompensated Fe spins associated with intermixed Fe atoms, at the interface with positive BFO FE surface charge. At the interface with negative FE surface charge, the ground state remains the configuration with compensated Fe spins, although the energy difference between the uncompensated and compensated Fe-spins configurations decreases drastically compared to the case of the chemically abrupt interface. The trends we find with intermixing are rationalised  based on a model description of the energetics of the exchange interactions.

Our study thus indicates that uncompensated Fe spins can emerge at the LSMO/BFO(001) interface as a consequence of cation Fe/Mn intermixing. Moreover, the switchable Fe FM state we find in our {\it ab initio} calculation, which depends on the sign of the BFO FE polarisation, may also account for the observed electric-field control of the EB recently reported in LSMO/BFO heterojunctions. We would like to note, however, that we do not exclude that more generally defects or incoherence in the octahedral tilting pattern in the proximity of the interface may be another possible source of uncompensated Fe spins.
\acknowledgments
The authors acknowledge support by the funding agencies CNPq and FAPERJ and by the ICTP/IAEA STEP programme. Calculations were performed on the SP6 cluster at the CINECA computer facility.

\begin{thebibliography}{10}
\expandafter\ifx\csname url\endcsname\relax\def\url#1{\texttt{#1}}\fi

\bibitem{ramesh2007multiferroics}
\Name{Ramesh R. \and Spaldin N.~A.} \REVIEW{Nature Materials }{6}{2007}{21}.

\bibitem{bibes2008spintronics}
\Name{Bibes M. \and Barth\'el\'emy A.} \REVIEW{Nature }{7}{2008}{425}.

\bibitem{wu2010reversible}
\Name{Wu S.~M. \etal} \REVIEW{Nature Materials }{9}{2010}{756}.

\bibitem{chu2008fmcontrol}
\Name{Chu Y.-H. \etal} \REVIEW{Nature }{7}{2008}{478}.

\bibitem{wang2003epitaxial}
\Name{Wang J. \etal} \REVIEW{Science }{299}{2003}{1719}.

\bibitem{zhao2006afcontrol}
\Name{Zhao T. \etal} \REVIEW{Nature Materials }{5}{2006}{823}.

\bibitem{krishnan2011microstructural}
\Name{Krishnan R. \etal} \REVIEW{Journal of Applied Physics
  }{109}{2011}{034103}.

\bibitem{yu2010interface}
\Name{Yu P. \etal} \REVIEW{Physical Review Letters }{105}{2010}{27201}.

\bibitem{borisevich2010suppression}
\Name{Borisevich A.~Y. \etal} \REVIEW{Physical Review Letters
  }{105}{2010}{87204}.

\bibitem{calderon2011magnetoelectric}
\Name{Calder\'on M.~J. \etal} \REVIEW{Physical Review B }{84}{2011}{024422}.

\bibitem{yu2012control}
\Name{Yu P. \etal} \REVIEW{PNAS }{109}{2012}{9710}.

\bibitem{QE-2009}
\Name{Giannozzi P. \etal} \REVIEW{Journal of Physics: Condensed Matter
  }{21}{2009}{395502}.
\newline\url{http://www.quantum-espresso.org}

\bibitem{neaton2005bulkBFO}
\Name{Neaton J.~B. \etal} \REVIEW{Physical Review B }{71}{2005}{014113}.

\bibitem{vailionis2011misfit}
\Name{Vailionis A. \etal} \REVIEW{Physical Review B }{83}{2011}{064101}.

\bibitem{junquera2003critical}
\Name{Junquera J. \and Ghosez P.} \REVIEW{Nature }{422}{2003}{506}.

\bibitem{he2010control}
\Name{He J. \etal} \REVIEW{Physical Review Letters }{105}{2010}{227203}.

\bibitem{pailloux2002nanoscale}
\Name{Pailloux F. \etal} \REVIEW{Physical Review B }{66}{2002}{14417}.

\bibitem{herger2008structure}
\Name{Herger R. \etal} \REVIEW{Physical Review B }{77}{2008}{085401}.

\end{thebibliography}
\end{document}